\documentclass[10pt]{article}
\usepackage{amsmath,amssymb}
\usepackage{bm}
\usepackage{color}
\usepackage[pdftex]{graphicx}
\usepackage{ascmac}
\usepackage{amsmath}
\usepackage{braket}
\usepackage{array}
\usepackage{float}
\usepackage[top=30truemm,bottom=35truemm,left=30truemm,right=30truemm]{geometry}
\usepackage{authblk}
\usepackage{comment}
\usepackage{url}
\title{\bf{Study of fermion pair events\\ at the 250 GeV ILC} }
\author[1]{Yuto Deguchi\thanks{Presenter. Talk presented at the International Workshop on Future Linear Collider (LCWS2018), Arlington, Texas, 22-26 October 2018. C18-10-22.}}
\author[1]{Hiroaki Yamashiro}
\author[1]{Taikan Suehara} 
\author[2]{Tamaki Yoshioka}
\author[3]{\\ Keisuke Fujii}
\author[1]{Kiyotomo Kawagoe}
\affil[1]{Department of Physics, Faculty of Science, Kyushu University, Fukuoka, Japan}
\affil[2]{Research Center for Advanced Particle Physics, Kyushu University, Fukuoka, Japan}
\affil[3]{High Energy Accelerator Research Organization (KEK), Tsukuba, Japan}
\date{}

\begin{document}
\maketitle

\begin{abstract}
Precise measurements of electroweak processes at the International Linear Collider (ILC) will provide unique opportunities to explore new physics beyond the Standard Model. Fermion pair production events are sensitive to new interactions involving a new heavy gauge boson or an electroweak interacting massive particle (EWIMP).We studied the mass reach of new particles at the ILC with $\sqrt{s}=250$ GeV by using $e^+ e^-\to e^+ e^-$and $e^+ e^-\to \mu^+ \mu^-$ events. We show that a mass reach for BSM particles can be determined with 90\% confidence level using a toy Monte Carlo technique.
\end{abstract}

\section{Introduction}
The International Linear Collider (ILC)\cite{design} is an electron-positron linear collider, which is planned to be constructed in Japan. The main purposes of the ILC are discoveries of new particles and precise measurement of the Higgs boson and the top quark. The center of mass energy will be 250 GeV with 20 km tunnel length. In the future, the center of mass energy can be upgraded to 1 TeV with 50 km tunnel length. \par
We know that the reaction of $e^+ e^-\to 2f$ events involve the emission and absorption or exchange of Z bosons in the Standard Model. In some models beyond the Standard Model (BSM), additional particles can also contribute to the interaction. Precise measurements of electroweak processes at the ILC can search these particles. \par
In this paper, We studied effects of new heavy gauge bosons ($Z'$ particles) with five BSM models and electroweak WIMPs with three BSM models.We calculated the mass reach of new particles at the ILC with $\sqrt{s}=250$ GeV by using $e^+ e^-\to e^+ e^-$and $e^+ e^-\to \mu^+ \mu^-$ events. We observed the effect of BSM models on the cross section of the final state leptons and the angular distributions. We calculated the mass limit of each model with toy Monte Carlo technique to make pseudo experiments.   

\section{Models}
In this study, we considered five types of $Z'$, which are Sequential Standard Model (SSM), $E_6$ group ($\chi$ model, $\psi$ model, $\eta$ model), and Alternative Left-Right symmetry (ALR)\cite{zprime}. These $Z'$ bosons are exchanged between leptons, like $Z$ boson, when electroweak interactions occur (See Fig.1). \par
We also considered three EWIMP models, which are Minimal Dark Matter (MDM), Higgsino, and Wino\cite{ewimp}.  These particles couple to the $Z'$ boson in a loop contribution (See Fig.2), which affects the $e^+ e^-\to 2f$ interaction. The quantum numbers of each EWIMP models are shown in Table 1.
\begin{table}[h]
 \begin{center}
 \caption{Quantum numbers of each model}
  \begin{tabular}{|c|c|c|} \hline
  model & $SU(2)_L$ & $U(1)_\gamma$ hypercharge \\ \hline
  MDM & $n=5$ (pentet) & 0 \\ \hline
  Higgsino & $n=2$ (doublet) & $\pm1/2$ \\ \hline
  Wino & $n=3$ (triplet) & 0 \\ \hline
  \end{tabular}
 \end{center}
 \end{table}

\begin{figure}[h]
  \begin{tabular}{c}
   \begin{minipage}{0.5\hsize}
    \includegraphics[width=7 cm]{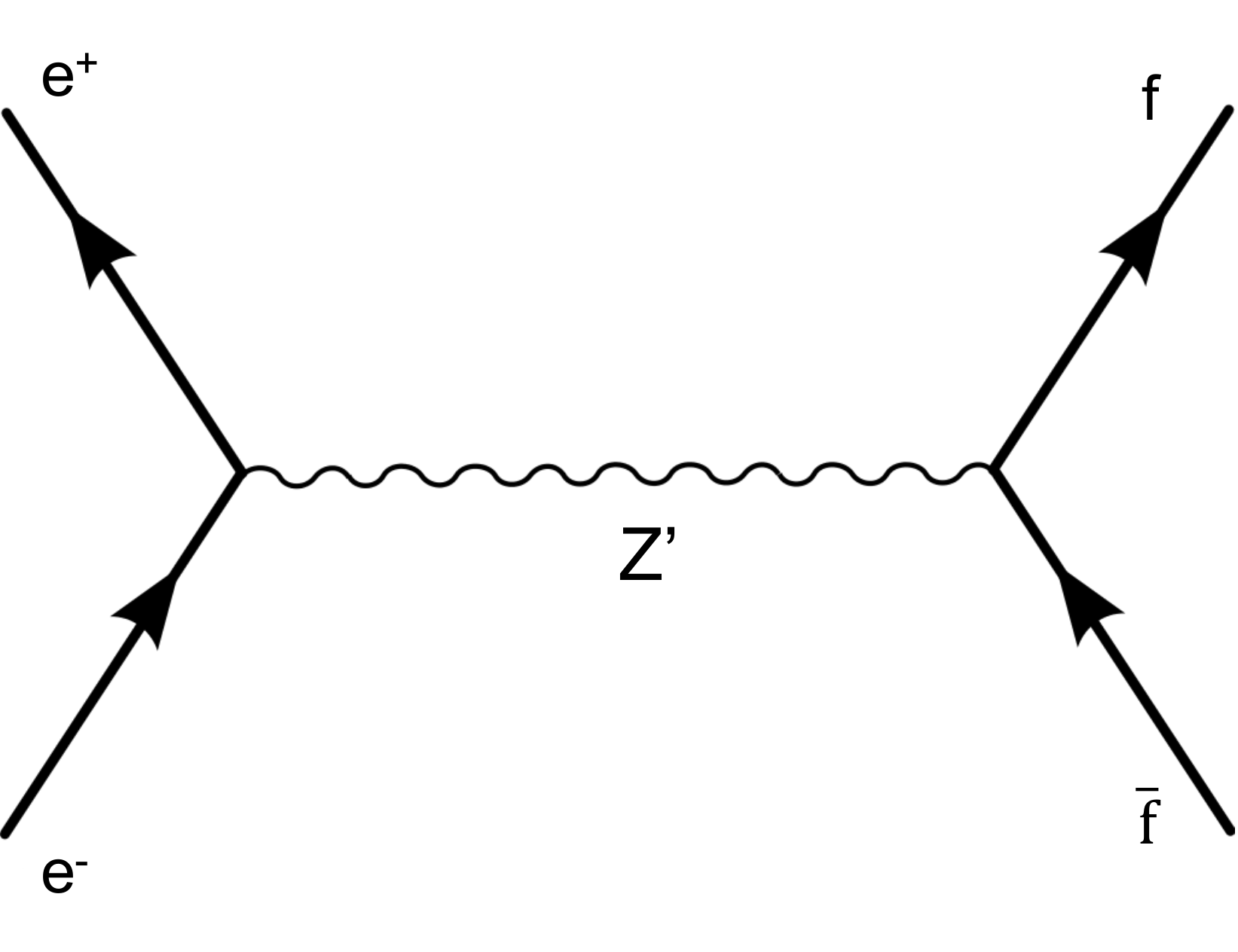}
    \caption{Feynman diagram involving $Z'$}
   \end{minipage}
  \end{tabular}
  \begin{tabular}{c}
   \begin{minipage}{0.5\hsize}
    \includegraphics[width=7 cm]{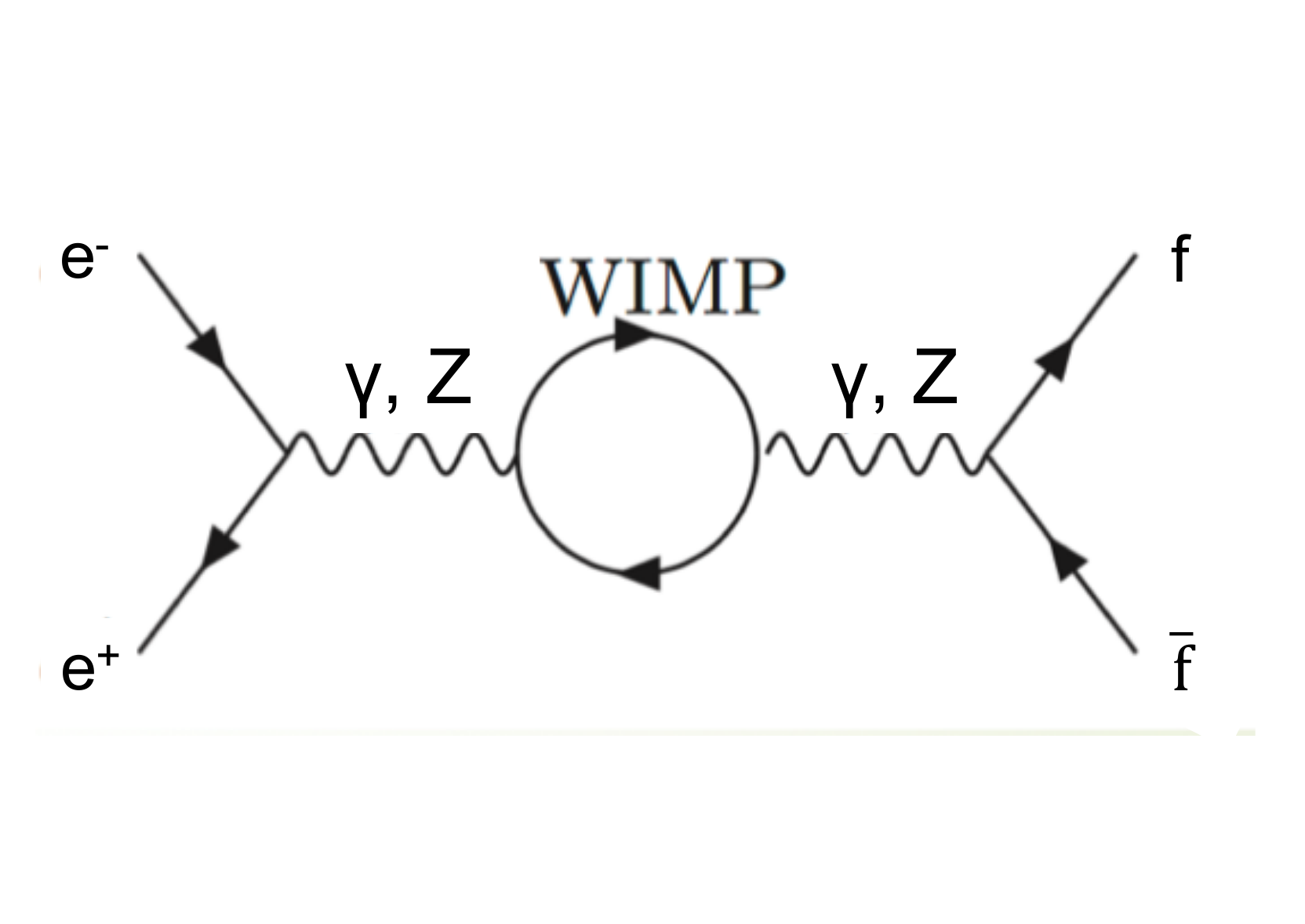}
    \caption{Feynman diagram involving EWIMP}
   \end{minipage}
  \end{tabular}
\end{figure}

\section{Simulation}
In this study, we used the ILCsoft\cite{ilcsoft} v01-16-p10 for the simulation and reconstruction. We used samples prepared for Detailed Baseline Design (DBD) of International Large Detector (ILD)\cite{ild} concept. The samples were generated by WHIZARD 1.95\cite{whizard}, and simulated by Mokka, with detector geometry of ILD\_v1\_o5. There is a special treatment of $e^+ e^-\to e^+ e^-$ samples, which includes preselection of xxxxxx. It was also prepared with weighted production, which produces the same amount of events for each $\cos\theta$ bins of 0.1 width. The ILD detector consists of vertex detector, a tracking detector, an electromagnetic calorimeter, and a hadron calorimeter. These detectors are located inside a superconducting solenoid with 3.5 Tesla magnetic field. The magnetic flux of the coil is returned in an iron yoke, equipped with a muon detector.The center of mass energy is $\sqrt{s}=250$ GeV and the total luminosity is 2000 ${\rm fb}^{-1}$. The beam polarization was set to 80\% left-handed for the electron beam and 30\% right-handed for the positron beam. The luminosity of left-handed electron and right-handed positron is 900 ${\rm fb}^{-1}$, and right-handed electron and left-handed positron is also 900 ${\rm fb}^{-1}$. After the detector simulation, the events are reconstructed by Marlin\cite{marlin}, including Pandora PFA\cite{pandora} for track-cluster matching.
 
\section{Event selection}
In this section, we explain the separation of background events from signal events.
\subsection{$e^+ e^-\to e^+ e^-$} 
To select the $e^+ e^-\to e^+ e^-$ events, we considered five cut terms. The final states of background events are  $e^+ e^-\to \mu^+ \mu^-$, $e^+ e^-\to \tau^+ \tau^-$, and $e^+ e^-\to 4f$ events, with only lepton in the final states. 
Firstly, we selected tracks with energy $E_{tr}>15$ GeV. Events which do not have positive and negative tracks meeting the requirement are removed. If there are more than one pair of such tracks, the track of highest energies are selected for both positive and negative charge. After that, we applied following selection conditions. 
\begin{itemize}
 \item The ratio of energy deposit at calorimeter to the energy of track : $E_{{\rm cal}}/E_{{\rm tr}}>0.8$ to reject muon events.
 \item $E_{{\rm ecal}}/(E_{{\rm ecal}}+E_{{\rm hcal}})>0.85$ to separate hadrons and muons. $E_{{\rm ecal}}$ is the energy deposit at the electromagnetic calorimeter, and $E_{{\rm hcal}}$ is the energy deposit at the hadron calorimeter.
 \item The energy sum of two tracks : $E_{{\rm sum}}>230$ GeV to exclude 4 fermion events and tau events. 
 \item $|\cos\theta|<0.95$ to cut t-channel  $e^+ e^-\to e^+ e^-$ events which have huge cross section at the forward angle. $\theta$ is the angle of the track with respect to the beam axis.
\end{itemize} 
 
The result of event selection is shown in Table 2. Most of background events can be excluded by these cut terms.
\begin{table}[h]
 \begin{center}
 \caption{Event selection of electron channel}
  \begin{tabular}{|c||r|r|r|r|} \hline 
  \multicolumn{1}{|c||}{Cut terms} & \multicolumn{1}{c|}{signal($e^+e^-$)} & \multicolumn{1}{c|}{$\mu^+\mu^-$} & \multicolumn{1}{c|}{$\tau^+\tau^-$} & \multicolumn{1}{c|}{4$f$} \\ \hline \hline
  all events & $8.26\times10^{8}$ & $9.73\times10^{6}$ & $1.17\times10^{7}$ & $7.73\times10^{6}$ \\ \hline
  $E_{{\rm tr}}>15$ GeV & $8.18\times10^{8}$ & $9.36\times10^{6}$ & $1.17\times10^{7}$ & $6.39\times10^{6}$ \\ \hline
  $E_{{\rm cal}}/E_{{\rm tr}}>0.8$ & $8.18\times10^{8}$ & $2.91\times10^{4}$ & $1.17\times10^{7}$ & $2.60\times10^{6}$ \\ \hline
  $E_{{\rm ecal}}/(E_{{\rm ecal}}+E_{{\rm hcal}})>0.85$ & $8.17\times10^{8}$ & $1.22\times10^{4}$ & $1.17\times10^{7}$ & $1.66\times10^{6}$ \\ \hline
  $E_{{\rm sum}}>230$ GeV & $3.68\times10^{8}$ & 14 & 21 & $1.14\times10^{4}$ \\ \hline
  $|\cos\theta|<0.95$ & $2.57\times10^{7}$ & 14 & 21 & $1.14\times10^{4}$ \\ \hline 
  \end{tabular}
 \end{center}
\end{table}

 \subsection{$e^+ e^-\to \mu^+ \mu^-$}
We selected most energetic tracks of positive and negative charge with $E_{{\rm tr}}>15$ GeV as same as $e^+ e^-\to e^+ e^-$ case, then we applied following selection conditions. 
\begin{itemize}
 \item $E_{{\rm cal}}/E_{{\rm tr}}<0.3$ and $E_{{\rm ecal}}/(E_{{\rm ecal}}+E_{{\rm hcal}})<0.45$ to reject electrons and hadrons.
 \item $E_{{\rm sum}}>230$ GeV to exclude 4 fermion and tau events. 
 \item $|\cos\theta|<0.95$ to cut t-channel  $e^+ e^-\to e^+ e^-$ events.
\end{itemize}

\begin{table}[h]
 \begin{center}
 \caption{Event selection of muon channel}
  \begin{tabular}{|c||r|r|r|r|} \hline 
  \multicolumn{1}{|c||}{Cut terms} & \multicolumn{1}{c|}{signal($\mu^+\,u^-$)} & \multicolumn{1}{c|}{$e^+e^-$} & \multicolumn{1}{c|}{$\tau^+\tau^-$} & \multicolumn{1}{c|}{4$f$} \\ \hline \hline
  all events & $9.73\times10^{6}$ & $8.26\times10^{8}$ & $1.17\times10^{7}$ & $7.73\times10^{6}$ \\ \hline
  $E_{{\rm tr}}>15$ GeV & $9.36\times10^{6}$ & $8.18\times10^{8}$ & $1.17\times10^{7}$ & $6.39\times10^{6}$ \\ \hline
  $E_{{\rm cal}}/E_{{\rm tr}}<0.3$ & $9.26\times10^{6}$ & $1.49\times10^{6}$ & $8.97\times10^{5}$ & $3.35\times10^{6}$ \\ \hline
  $E_{{\rm ecal}}/(E_{{\rm ecal}}+E_{{\rm hcal}})<0.45$ & $2.91\times10^{6}$ & $1.49\times10^{6}$ & $7.80\times10^{5}$ & $1.33\times10^{6}$ \\ \hline
  $E_{{\rm sum}}>230$ GeV & $7.49\times10^{5}$ & 56 & $4.52\times10^{3}$ & 32 \\ \hline
  $|\cos\theta|<0.95$ & $7.48\times10^{5}$ & 56 & $4.48\times10^{3}$ & 32 \\ \hline 
  \end{tabular}
 \end{center}
\end{table}
The result of event selection is shown in Table 3. We could also exclude most of background events.

\section{Analysis}
After the event selection, we obtained the angular distribution for each channel with 20 bins, shown in Fig.3 and Fig.4. Vertical axis of the electron channel is logarithmic scale, and vertical axis of the muon channel is linear scale. The mass reach of new particle were calculated using these angular distributions. \par 
\begin{figure}[t]
 \begin{tabular}{c}
  \begin{minipage}{0.5\hsize}
   \includegraphics[width=7 cm, clip]{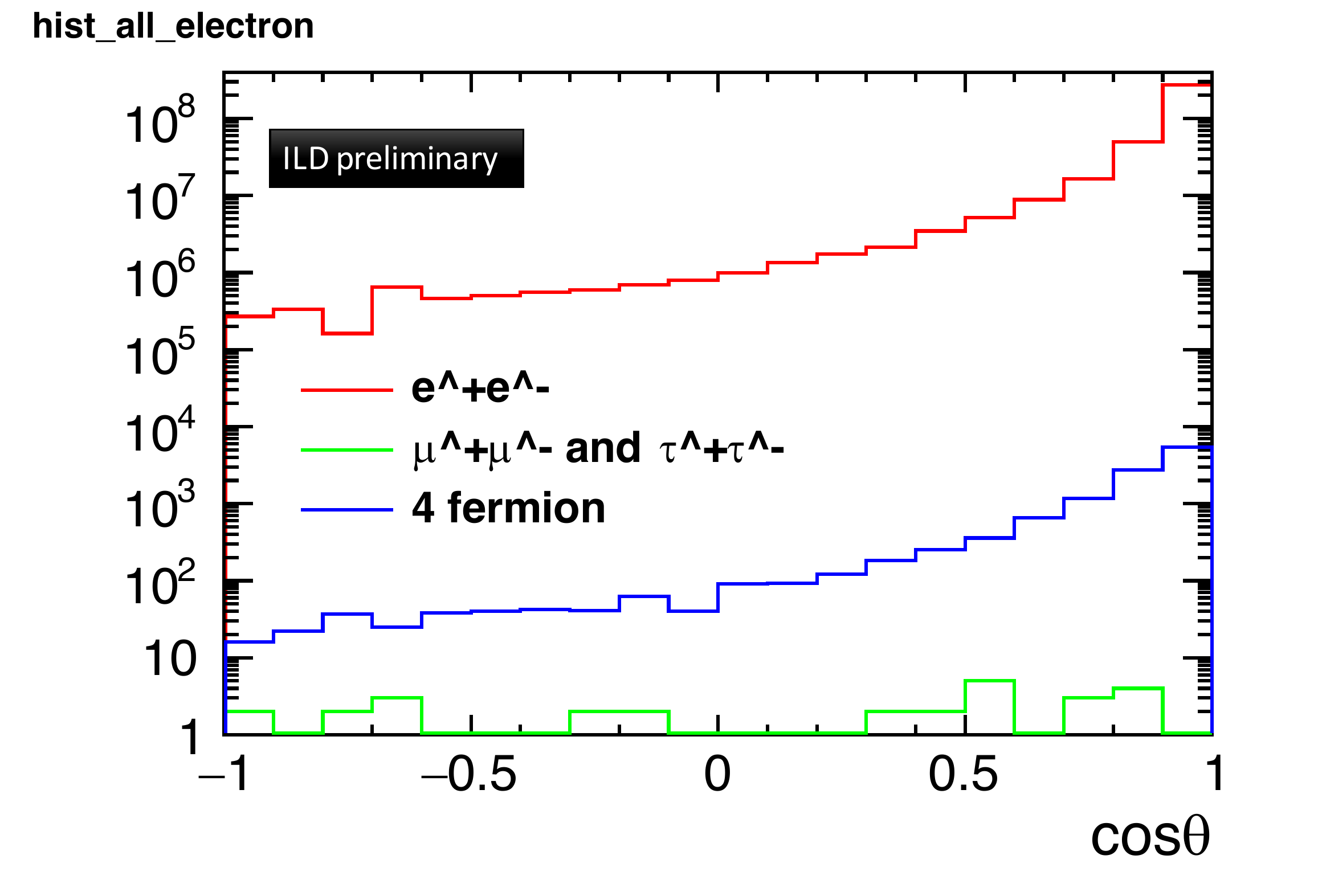}
   \caption{Angular distribution of electron channel}
  \end{minipage}	
 \end{tabular}
 \begin{tabular}{c}
  \begin{minipage}{0.5\hsize}
   \includegraphics[width=7 cm, clip]{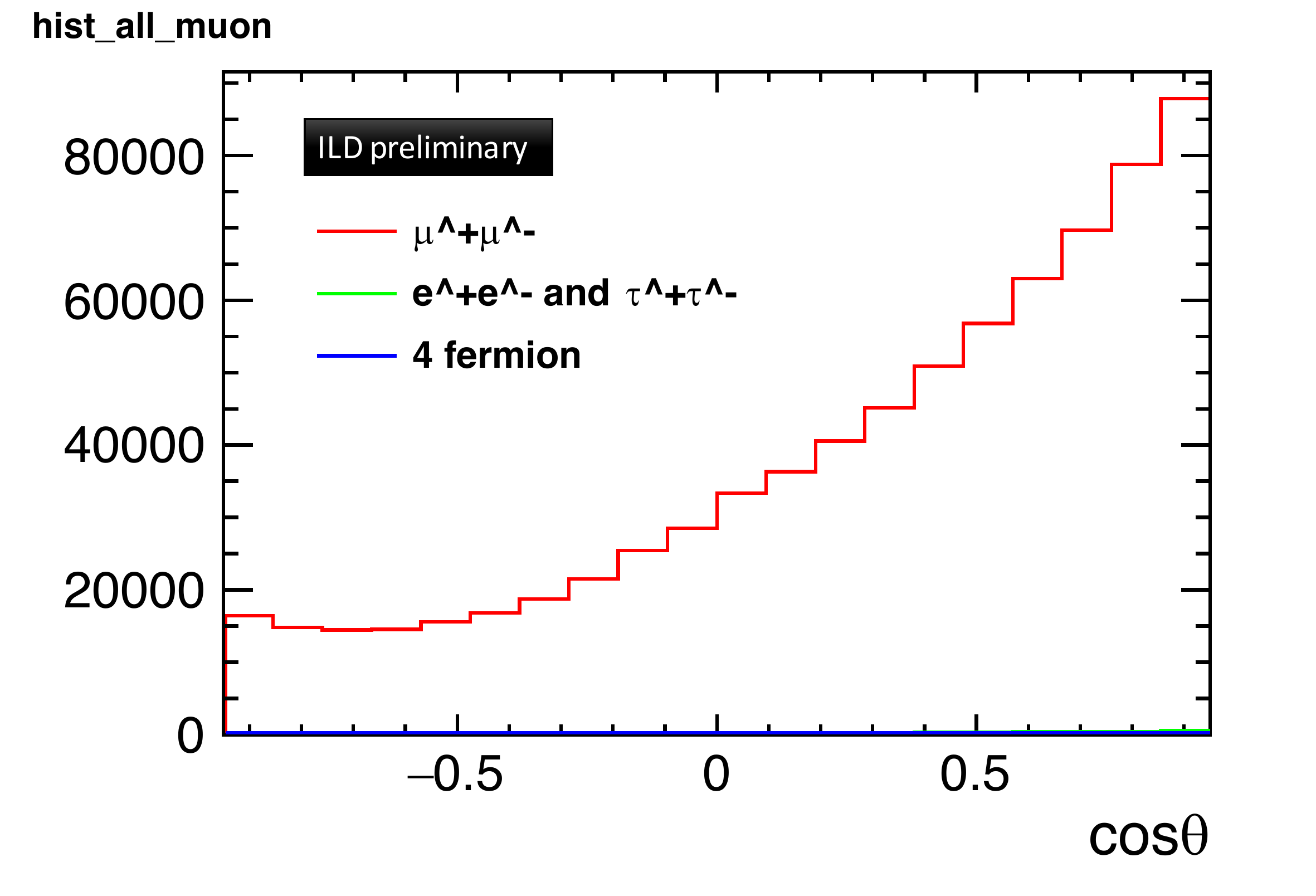}
   \caption{Angular distribution of muon channel}
  \end{minipage}
 \end{tabular}
\end{figure}
First, we calculated deviations of particle cross section with each BSM model at each bin. Second, we made pseudo experiments for each bin using SM angular distribution with fluctuation of the Poisson distribution. Third, we calculated $\chi^2$ between the pseudo experiments and the each BSM model to obtain probability.  At this time, $0.1$\% systematics are assumed for each bin. The distributions of probability are shown in Fig.5 and Fig.6. Finally, We obtained medians from the probability distributions of each BSM model. Mass dependence of the probability for each particles is shown in Fig.7 and Fig.9. The blue lines where the probability is $0.1$ show $90$\% confidence level mass limit. These values for each particle are shown in Table 4 and Table 5. 

\begin{figure}[b]
 \begin{center}
  \includegraphics[width=16cm, clip]{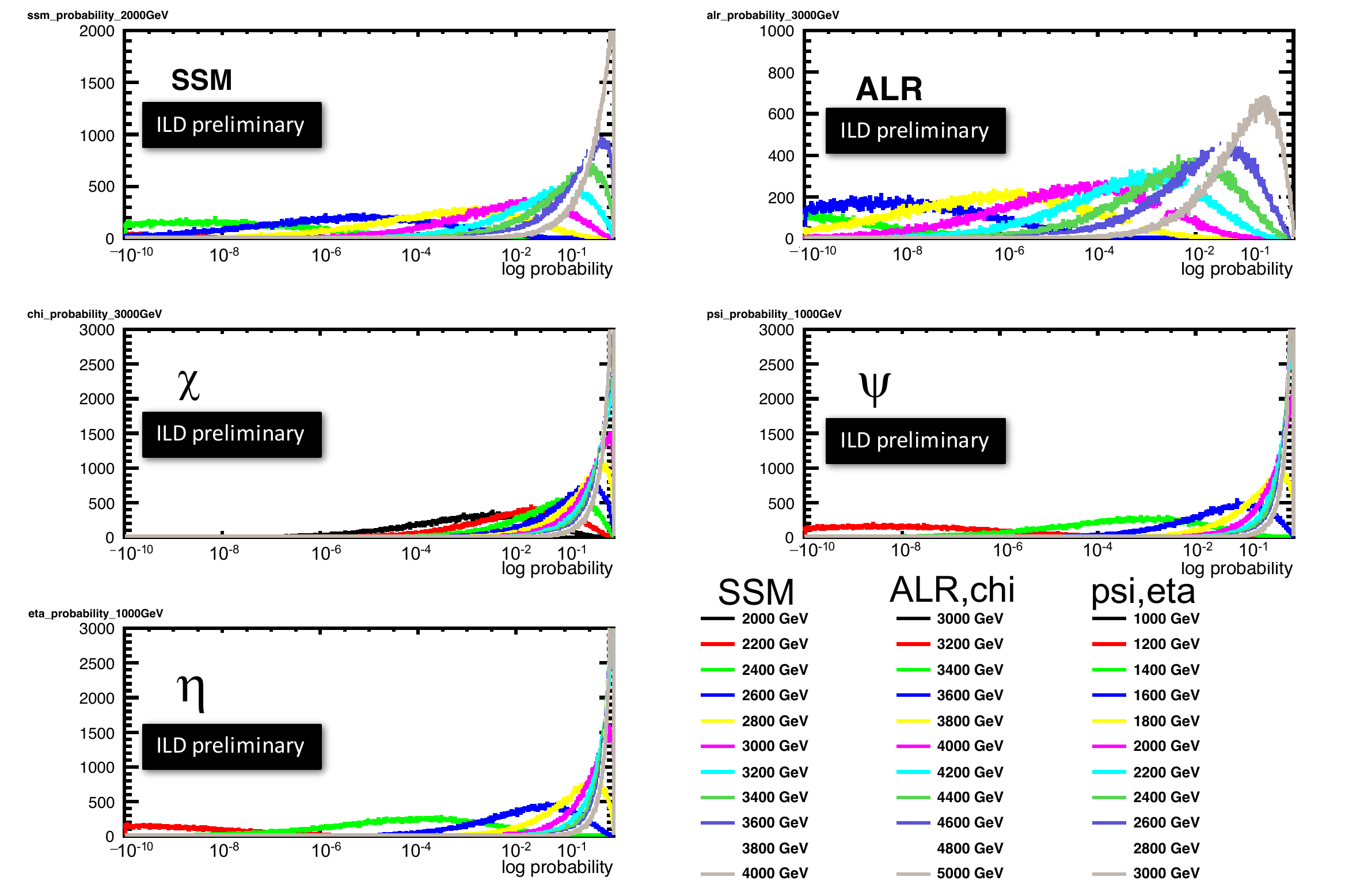}
  \caption[Probability distribution of $Z'$]{Probability distribution of $Z'$ models with 100,000 pseudo experiments. The colors show different masses of $Z'$ bosons as shown in the legend.}
 \end{center}
\end{figure}

\begin{figure}[b]
 \begin{center}
  \includegraphics[width=16 cm, clip]{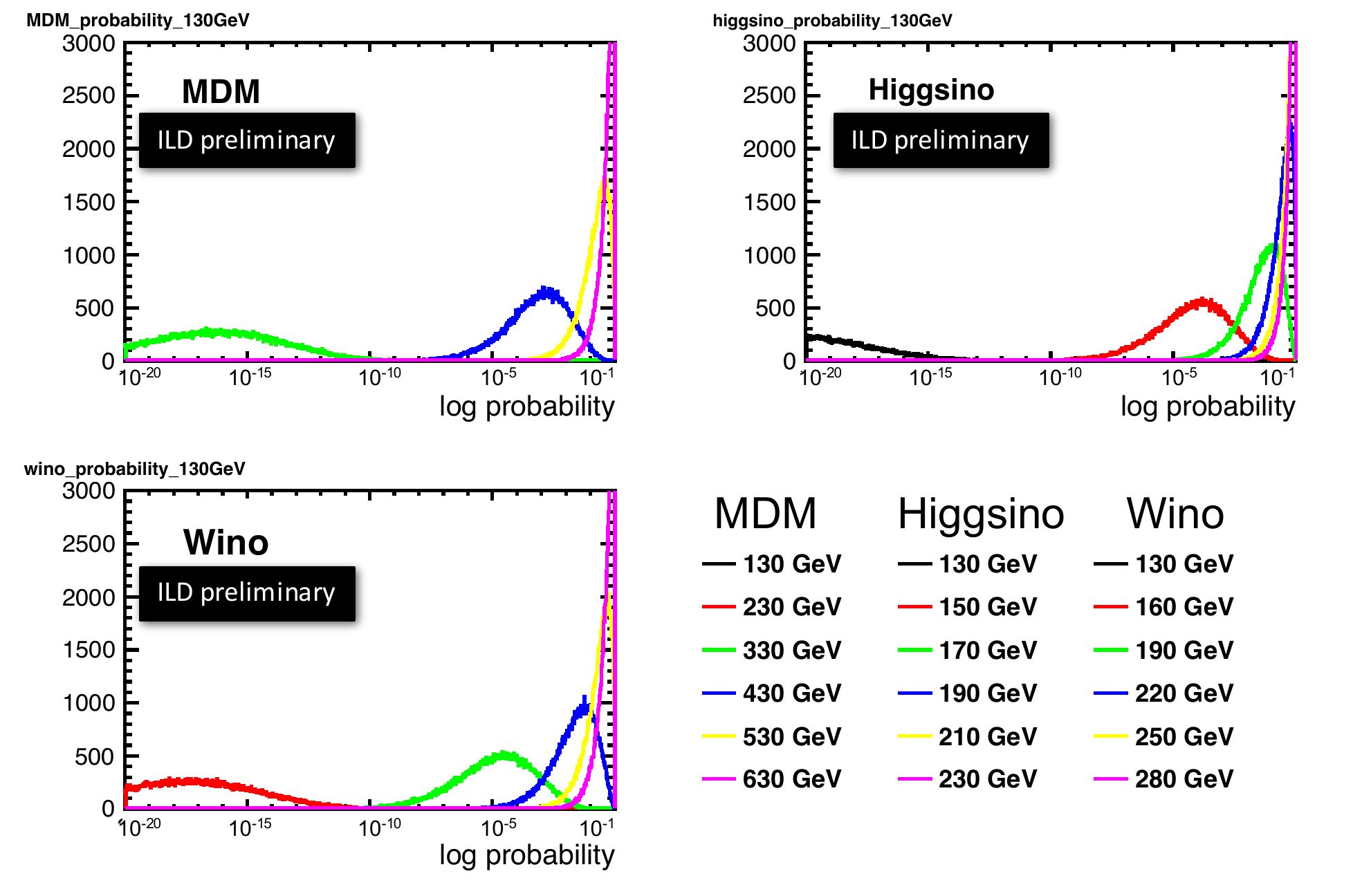}
  \caption[Probability distribution of EWIMP]{Probability distribution of EWIMP models with 100,000 pseudo experiments. The colors show different masses of EWIMPs as shown in the legend.}
 \end{center}
\end{figure}

\begin{figure}[b]
 \begin{center}
    \includegraphics[width=16 cm, clip]{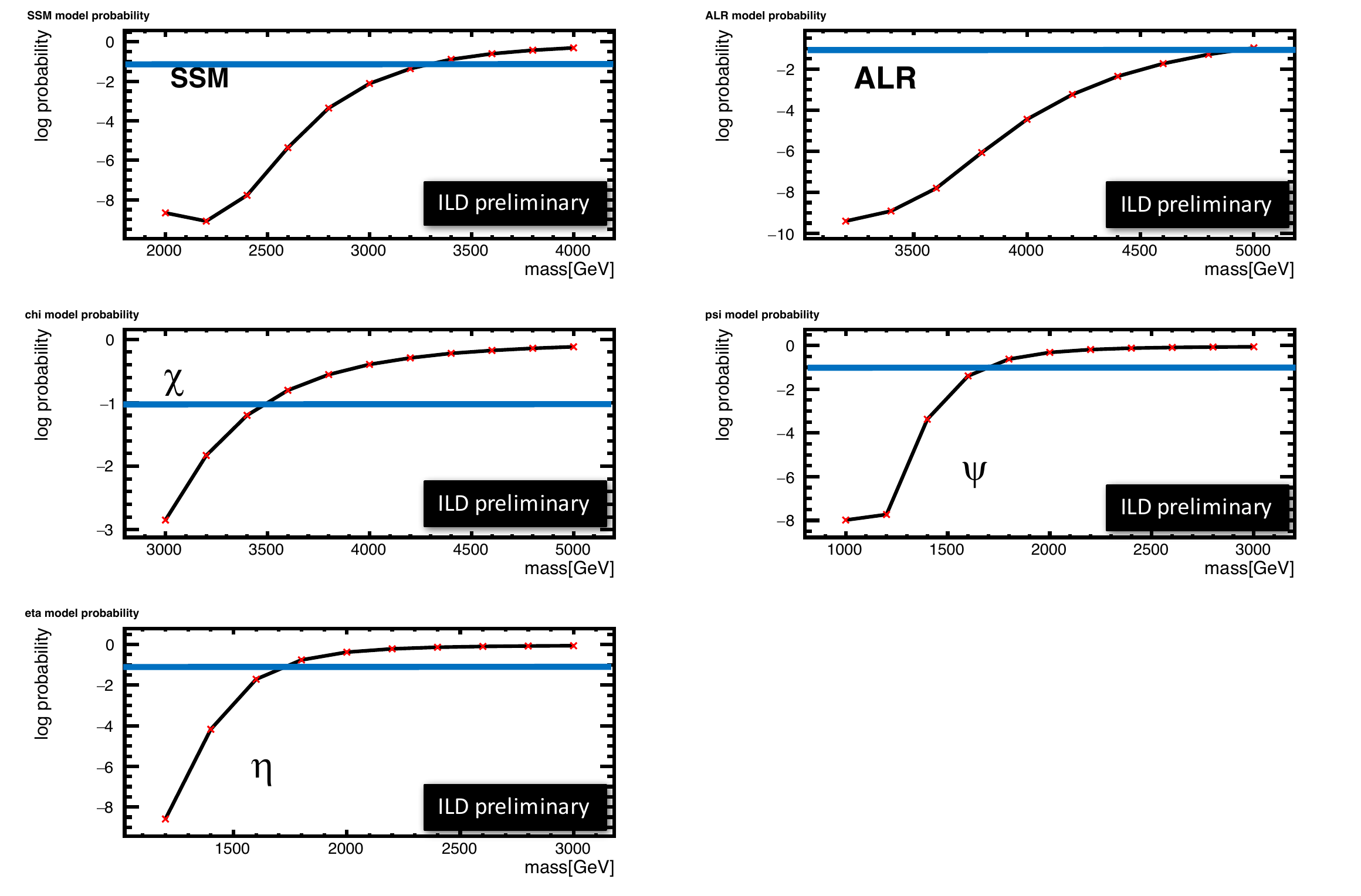}
    \caption[Median of the probability distribution obtained by pseudo experiments with $Z'$ models]{Median of the probability distribution obtained by pseudo experiments with $Z'$ models. The blue lines show exclusion limit at 90\% confidence level.}
 \end{center}
\end{figure}

\begin{figure}[b]
 \begin{center}
    \includegraphics[width=16 cm, clip]{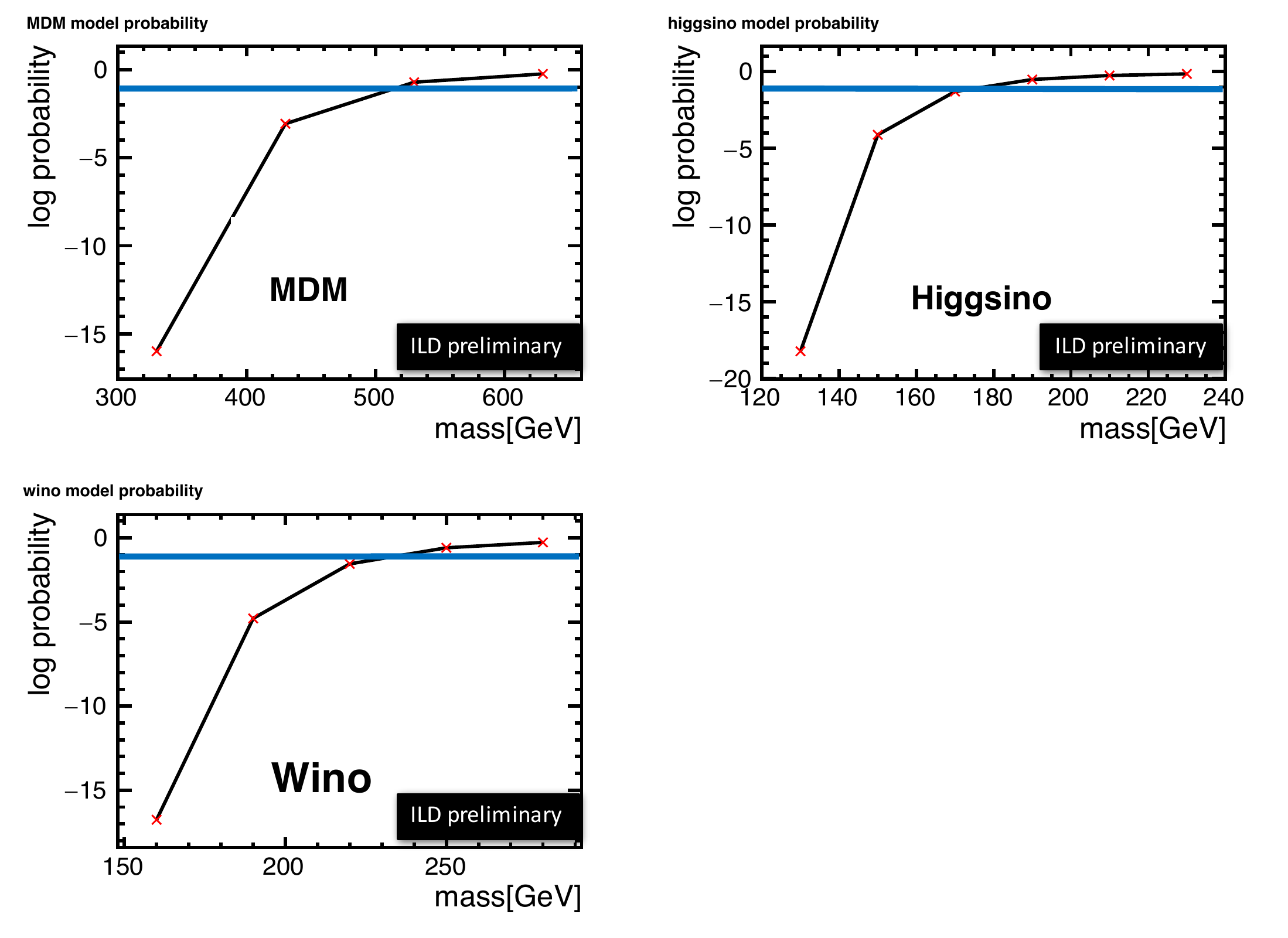}
    \caption[Median of the probability distribution obtained by pseudo experiments with EWIMP models]{Median of the probability distribution obtained by pseudo experiments with EWIMP models. The blue lines show exclusion limit at 90\% confidence level.}
 \end{center}
\end{figure}

\begin{table}[h]
 \begin{center}
  \begin{tabular}{c}
   \begin{minipage}{0.5\hsize}
    \begin{center} 
     \caption{Mass reach of $Z'$}
     \begin{tabular}{| l | r |} \hline 
     BSM & 
     \begin{tabular}{ l }
     mass reach (90\% CL) 
     \end{tabular} \\ \hline \hline
     SSM & 3.4 TeV \\ \hline
     ALR & 5.0 TeV \\ \hline
     $\chi$ & 3.5 TeV \\ \hline
     $\psi$ & 1.7 TeV \\ \hline
     $\eta$ & 1.7 TeV \\\hline
     \end{tabular}
    \end{center}
   \end{minipage}
   
   \begin{minipage}{0.5\hsize}
    \begin{center} 
     \caption{Mass reach of EWIMP}
     \begin{tabular}{| l | r |} \hline 
     BSM & 
     \begin{tabular}{ l }
     mass reach (90\% CL) 
     \end{tabular} \\ \hline \hline
     MDM & 500 GeV \\ \hline
     Higgsino & 180 GeV \\ \hline
     Wino & 240 GeV \\ \hline
     \end{tabular}
    \end{center}
   \end{minipage}
  \end{tabular}
 \end{center}
\end{table}

The mass reaches of $Z'$ are comparable to LHC. Thus we have to conclude that the CM energy should be much higher than 250 GeV to obtain significant gain of measurement from LHC results. The mass reaches of EWIMP are bigger than 125 GeV, which is the limited direct search, so we think the $e^+ e^-\to 2f$ can be a powerful prove to look for such a new physics.
\section{Summary}
We are investigating the effects from BSM models on $e^+ e^-\to 2f$ final states at $e^+ e^-$ colliders. We obtained the mass reach of the new BSM particles at 90\% confidence level at ILC with $\sqrt{s}=250$ GeV. In this paper, we considered the mass reach using $e^+ e^-\to e^+ e^-$ and $e^+ e^-\to \mu^+ \mu^-$, but we didn't include $e^+ e^-\to \tau^+ \tau^-$. We expect that better mass reach can be obtained by including tau and hadron channel. 

\section*{Acknowledgements}
We thank ILD physics and software group for generating events and many support, and we thank Dr. Satoshi Shirai for providing the cross sections with EWIMP models. This work was supported by JSPS KAKENHI Grant Number 16H02176.


\end{document}